\documentclass[aps,prl,showpacs,twocolumn,amsmath,amssymb,10pt]{revtex4-1}
\usepackage{graphicx}
\usepackage{color}

\newcommand{\dd}{\text{d}}      % for integral/derivative
\newcommand{\mean}[1]{\left\langle #1 \right\rangle} % average
\def\l{\lambda}

\def\beq{\begin{equation}}
\def\ee{\end{equation}}
\def\bi{\begin {itemize}}
\def\ei{\end{itemize}}

\def\l{\lambda}

\def\njp{{n_j}^+}
\def\njm{{n_j}^-}

\def\tot{_{\rm tot}}

\def\beq{\begin{equation}}
\def\ee{\end{equation}}

\def\bi{\begin {itemize}}
\def\ei{\end{itemize}}

\def\wmn{w_{mn}}
\def\wnm{w_{nm}}

\def\ex{^{\rm ex}}
\def\hk{^{\rm hk}}
\def\tot{^{\rm tot}}
\def\sys{^{\rm sys}}

\def\out{^{\rm out}}
\def\in{^{\rm in}}

\def\I{{\cal I}}
\def\i{\text{\scriptsize $\cal{I}$}}

\begin{document}

\title{Thermodynamics of genuine non-equilibrium states under feedback control}
\author{David Abreu and Udo Seifert}
\affiliation{{II.} Institut f\"ur Theoretische Physik, Universit\"at Stuttgart, 70550 Stuttgart, Germany}
\pacs{05.70.Ln, 05.40.-a}

\begin{abstract}
For genuine non-equilibrium states that even at fixed external control parameter exhibit dissipation, we extend the Hatano-Sasa equality to processes with feedback control. The resulting bound on the maximal extractable work is substantially sharper than what would follow from applying the Sagawa-Ueda equality to transitions involving such states. For repeated measurements at short enough intervals, the power thus extracted can even exceed the average cost of driving as demonstrated explicitly with a simple, analytically solvable example.
\end{abstract}

\maketitle

{\sl Introduction and main results.--} 
According to the second law of thermodynamics, the work $W\out$ extracted in any process applied to  a system coupled to a heat bath of temperature $T$ cannot exceed the free energy difference ($-\Delta F$) between the initial and final equilibrium state. If, however, additional information $\I$ about the state of the system during this process becomes available through measurements, more work can be extracted by adjusting a  control parameter driving the transition accordingly. For such processes with feedback, the thermodynamic bound   
\beq
W\out \leq -\Delta F + k_BT~ \I  ,
\label{eq1}
\ee
with $k_B$ Boltzmann's constant, has been proven for single and for repeated measurements both for Hamiltonian and stochastic dynamics \cite{kawa07,cao09,saga10,horo10,ponm10,saga11}. This relation has been verified in an experiment using a colloidal particle trapped by two feedback-controlled electric fields \cite{toya10a}. Theoretical case studies of Eq. (\ref{eq1}) include Szilard-type engines using two-state systems or a partitioning of phase-space in two regions \cite{mara10,vaik11} and Langevin systems \cite{kim04,fuji10}. For feedback-reversible processes, the bound in Eq. (\ref{eq1}) can be saturated \cite{horo11}. For finite-time processes, the maximum extractable work has been investigated for a Brownian particle in time-dependent harmonic traps \cite{abre11}. Characteristically, all the systems considered so far  reach genuine equilibrium for a constant external control parameter $\lambda$ with a Boltzmann-type distribution and zero currents.

On the other hand, there is a large class of systems  which at constant $\lambda$ approach a genuine non-equilibrium steady state (NESS) with permanent dissipation which under isothermal conditions is called the housekeeping heat $Q\hk=T\Delta S\hk$ \cite{hata01}. Examples include essentially all motor proteins since they typically move with an on average  constant speed  at fixed external conditions like  concentrations of chemicals or applied forces or torques (see, e.g., \cite{toya10,haya10} and refs. therein). Likewise, a colloidal particle driven by a constant force along a periodic potential can serve as another paradigm for such a NESS \cite{blic07,gome09}.

In this paper, we study the thermodynamics of feedback control of such non-equilibrium states. How much (more) work can be extracted if information acquired through measurement is used for optimal driving? Is it possible to extract net work that exceeds even the cost of driving from transitions between such states? One could be inclined to try to answer these questions naively by generalizing Eq. \eqref{eq1} to this situation in the form
\beq
- \Delta S\tot \leq k_B\I,
\label{eq2}
\ee
with $\Delta S\tot$ being the total entropy change of system and heat bath, thereby getting rid off the no longer well-defined free energy when dealing with genuine NESSs. Indeed, as shown towards the end of the paper, Eq. (\ref{eq2}) is true for such systems even though this fact seems not to have been stated yet explicitly. On the other hand, Eq. (\ref{eq2}) is about as useless as a sensitive bound for practical situations as the ordinary second law (Eq. (\ref{eq2}) with $\I=0$) is for non-feedback driven transitions between NESSs. The permanent dissipation required to drive these NESSs masks all finer details associated with the transition between such states. Addressing this issue for transitions without feedback, Hatano and Sasa \cite{hata01} have derived an equality from which they obtained a sensitive bound focussing on the extra dissipation.

As our first main result, we will generalize their result to feedback driven processes and derive the equality
\beq
\langle \exp [-(\Delta s\tot -\Delta s\hk+ \i )] \rangle =1
\label{eq3}
\ee 
for  trajectory-dependent quantities indicated throughout by the corresponding small letters and setting $k_B=T=1$ from now on such that heat can be replaced by the corresponding entropy change. The average is over arbitrary initial conditions and any number of measurements with subsequent adaption of a control parameter implementing the feedback. This generalization shares some technical analogies with Sagawa and Ueda's extension  \cite{saga10} of the Jarzynski relation \cite{jarz97} to feedback control for transitions between equilibrium states. Our work, however, applies to fundamentally different physical systems, such as artificial or biological motors driven by unbalanced chemical reactions, or basically any other transport problem, reaching a NESS at constant control parameter.

As second main result, for repeated measurements at regular intervals, Eq. \eqref{eq3} will be shown to imply
\beq
\dot W\out \leq \dot W\in -\dot Q\hk +\dot \I
\label{eq4}
\ee
as a bound for the power (which, as any rate, will be denoted by a dot) that can be extracted under isothermal conditions from such a feedback driven steady-state machine. Here $\dot W\in$ is the cost spent in driving the system. For a large enough rate of acquiring information, i.e. for $\dot \I > \dot Q\hk$, a net gain becomes possible since the extracted power may then exceed the cost of driving.
\begin{figure}
\centering
\includegraphics[width=0.7\columnwidth]{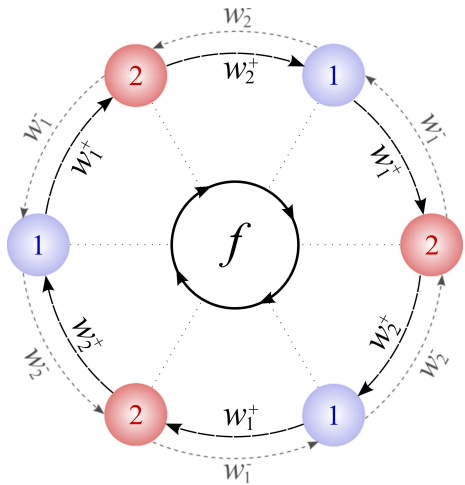}
\caption{Model rotary motor: The driving force $f>0$ induces $60^\circ$-clockwise rotations along which the internal states $1$ and $2$ alternate. The transition rates $w_{1,2}^{+,-}$ are given by \eqref{eq:w1} and \eqref{eq:w2}.}
\label{fig1}
\end{figure}

{\sl Model system.--}
We first illustrate this inequality for a paradigmatic example for which all relevant quantities can be calculated analytically. We consider a system which has two internal states labelled $1$ and $2$ that alternate in a spatial direction. The NESS is generated by a non-vanishing force $f>0$ that drives the system in a cyclic fashion from state 1 to 2 to 1 to 2 and so on. An example of such a system is given in Fig. \ref{fig1}, which is a simplified model for a rotary motor protein like the F1-ATPase \cite{haya10,toya10}. For each clockwise  transition determined by the rates $w_1^+$ and $w_2^+$, driving the system costs the work $f$ (setting for simplicity the distance between the states symmetrically to 1), leading to a non-zero current. Likewise, the same work is recovered for any thermally excited step in counter-clockwise direction with the rates $w_1^-$ and $w_2^-$. Additionally, we assume that the states 2 have an energy $E$ higher (or, for $-E<0$, lower) than the states 1 with energy $0$. $E$ will serve as the control parameter $\lambda$. The rates 
\beq
w_1^+=w_0^2/w_1^-=w_0\exp[(f-E)/2]
\label{eq:w1}
\ee
and 
\beq
w_2^+=w_0^2/w_2^-=w_0\exp[(f+E)/2]
\label{eq:w2}
\ee
are chosen such that they obey local detailed balance \cite{seif10} and become maximally symmetric to facilitate the analytical solution, $w_0$ being a characteristic frequency of the system set to $1$ in the following. Maintaining this NESS without any feedback interference is easily calculated to cost 
\beq
\dot W\in = \dot Q\hk =  2f \frac{\sinh(f/2)}{\cosh(E/2)}.
\label{eq:hk}
\ee
\begin{figure}
\centering
\includegraphics[width=0.8\columnwidth]{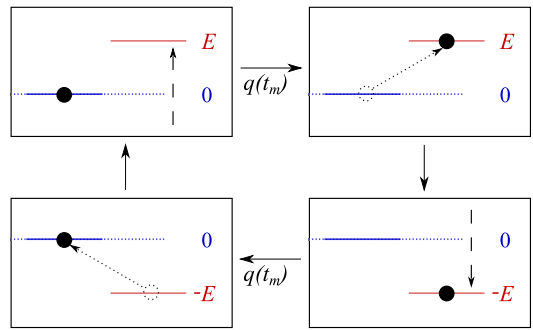} 
\caption{Feedback scheme: The energy of the internal state 2 is changed from $E>0$ to $-E$ if the system has jumped to this state of higher energy, which happens with probability $q(t_m)$. If state $2$ is empty, it can be moved to $E>0$. }
\label{fig2}
\end{figure}

The feedback scheme is implemented as follows, see Fig. \ref{fig2}. Perfect measurements at regular intervals separated by a time $t_m$ are supposed to reveal whether the system is in a state 1 or 2. If the system is in 1, we set the energy of state 2 to $E>0$ at no cost since the level is not occupied. If at the next measurement the system is still in 1 we do nothing. As soon as a subsequent measurement reveals that the system is in a state 2, we decrease the energy of this state to $-E<0$, thus extracting the work $W\out = 2E>0$. The next interval then starts with the system again in the lower energy state (which is now state $2$). As soon as a measurement reveals that the system has jumped to 1, we raise the energy of the now empty state 2 to $+E$ again at no cost. In the steady state finally reached in this scheme, the system is as likely found in 1 as in 2 on average. The probability $q(\tau)$ that it will have switched its state at time $\tau$ after the last measurement obeys the master equation
\beq
\dot{q}(\tau) = (w_1^++w_2^-)(1-q(\tau))-(w_2^++w_1^-)q(\tau),
\ee
the solution of which is
\beq
q(\tau)= \frac {\exp (-E/2)}{2 \cosh (E/2)} [1-\exp(-\gamma \tau)]
\ee
with the relaxation rate $\gamma\equiv w_1^+ + w_1^- + w_2^+ + w_2^-$.

The average power extracted from this machine in this feedback-driven NESS is 
\beq
\dot W\out(t_m) = E q(t_m)/t_m  ,
\ee
since work can be extracted only after a transition from $1$ to $2$. Likewise, the rate with which information is acquired in this binary measurement scheme becomes 
\beq
\dot \I(t_m)= -[q\ln q +(1-q)\ln(1-q)]/t_m > \dot W\out(t_m) 
\ee 
with $q\equiv q(t_m)$.

\begin{figure}
\centering
\includegraphics[width=0.95\columnwidth]{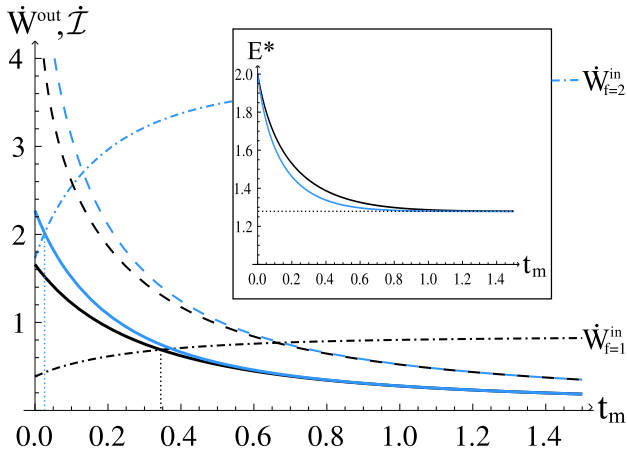}
\caption{Extracted power $\dot{W}\out$ (solid lines) as a function of the time $t_m$ between two measurements, maximized with respect to $E$ (see insert) for the values $f=1$ (black) and $f=2$ (light blue) of the driving force. The corresponding driving power $\dot{W}\in$ (dash-dotted) and the rate of acquiring information $\dot{\I}$ (dashed) are also represented. The two vertical dotted lines show the time $\hat t_m$ above which the driving requires more power than can be extracted, i.e. $\dot{W}^{out}(\hat t_m)=\dot{W}^{in}(\hat t_m)$.}
\label{fig3}
\end{figure}

Both rates are shown in Fig. \ref{fig3} as a function of the time $t_m$ between two measurements and for the energy $E^*(t_m,f)$ which at fixed $t_m$ and $f$ maximizes $\dot{W}\out$. For $t_m\gamma\ll 1$, the system hardly has had a chance to reach the higher level. Still, the optimal extracted power $\dot W\out= (4/e)\cosh(f/2)$, using $E^*(0,f)=2$, is the largest in this limit, in which the rate of acquiring  information  diverges. For $t_m\gamma\gg 1$, both $\dot{W}^{out}$ and $\dot{\I}$ tend to $0$ and the optimal extracted power is obtained for $E^*(\infty,f)\simeq 1.278$.

The process with feedback requires less power for driving than the one without feedback as given in (\ref{eq:hk}). The former's power can be determined from the average instantaneous cost of driving (counting the time $\tau$ from the last measurement) which is
\beq
\mathcal{\dot W}\in(\tau) = f [q(\tau)(w_2^+-w_1^-) + (1-q(\tau))(w_1^+ -w_2^-)].
\ee
The time averaged rate 
\beq
\dot W\in(t_m)\equiv\int_0^{t_m} \mathcal{\dot W}\in(\tau)d\tau/t_m=\dot Q\hk(t_m)
\ee
increases with $t_m$ as shown in Fig. \ref{fig3} and is equal to Eq. \eqref{eq:hk} for $t_m\to\infty$ since the system has then reached its steady-state.

In the limit $t_m\gamma\ll 1$, the extracted power exceeds the cost of driving for $E> f\tanh(f/2)$. Such an overall positive performance can persist up to a time $\hat t_m(E,f,\gamma)$ but, of course, not for arbitrary long intervals between measurements since the gain per measurement is bounded whereas the total cost of driving the system scales linearly with time.

{\sl Proof.--}  
The general results, Eqs. (\ref{eq3}) and (\ref{eq4}), will now be proven assuming a Markovian dynamics on a set of discrete states ${n}$. At time $\tau$, the system is in a state $n(\tau)$, jumping at discrete times $\tau_j$ from state $\njm$ to state $\njp$. A transition between state $m$ and state $n$ occurs with a rate $ \wmn(\l_j)$ which depends on the instantaneous value $\l_j\equiv \lambda(\tau_j)$ of a control parameter $\l(\tau)$. The solution of the master equation
\beq
\partial_\tau p(n,\tau) = \sum_m\left[\wmn(\l) p(m,\tau) -\wnm(\l)p(n,\tau)\right]
\label{eq:me}
\ee
for the probability $p(n,\tau)$ to find the system in state $n$ at time $\tau$ must be distinguished from the stationary solution $p_n^s(\l)$ corresponding to the instantaneous value of the control parameter.

Integrated along a trajectory, $0\leq \tau\leq t$,  the total entropy balance reads \cite{hata01,seif05a}
\beq
\Delta s\tot=\Delta s\sys + \Delta s\hk + \Delta s\ex .
\label{eq:ent}
\ee
Here, $\Delta s\sys\equiv - \ln p(n_t,t) + \ln p(n_0,0)$ is the change in system entropy with $n(0)\equiv n_0$ and $n(t)\equiv n_t$. The entropy change associated with the heat bath is the sum of housekeeping entropy $\Delta s\hk$ and excess entropy
\beq
\Delta s\ex = \int_0^t \dd\tau \sum_j\delta(\tau-\tau_j) \ln[p^s_{\njp}(\l_j)/p^s_{\njm}(\l_j)]
\ee
where the sum runs over all jumps in the corresponding time interval of length $t$.

For processes without feedback, this excess entropy obeys an integral fluctuation theorem
\beq
\left\langle \frac{g(n_t)}{p_0(n_0)}\exp(-\Delta s\ex)\right\rangle = 1
\label{eq:int1}
\ee
valid for any initial condition $p_0(n)$ and any normalized function $g(n)$ evaluated at the end point \cite{hata01,seif05a}. By choosing for $p_0$ and $g$ the stationary distribution corresponding to the initial and final value of the control parameter $\l$, respectively, this relation becomes precisely the Hatano-Sasa relation \cite{hata01}. For the following proof, we will, however, keep $p_0(n)$ as  an arbitrary initial condition. We choose $g(n)=\delta_{n,m}$ which leads, by using Bayes' theorem, to the conditioned average
\beq
\left\langle p_0^{-1}(n_0)\exp(-\Delta s\ex) |n_t=m)\right\rangle p(m,t) = 1
\label{eq:int2}
\ee
valid for any $m$ from which we get the equivalent
\beq
\sum_{n_0} \mean{\exp(-\Delta s\ex)|n_0,n_t=m}p(m,t)=1
\label{eq:int3}
\ee
by conditioning the average also on the initial state and thus cancelling the term $p_0^{-1}(n_0)$.

For the process with feedback, we first consider the case of two measurements at time $t_1$ and $t_2$ with outcomes $y_1$ and $y_2$, respectively. The trajectory-dependent information acquired with the first measurement is \cite{saga10,horo10,ponm10,saga11}
\beq
\i_1= \ln[p_1(n_1,t_1|y_1)/p(n_1,t_1)]
\label{eq:i1}
\ee
where $p_1(n_1,t_1|y_1)$ is the probability to find the system in state $n_1\equiv n(t_1)$ given the measured value $y_1$. This function now serves as the initial condition for the further time evolution of $p_1(n,\tau|y_1)$ which obeys the master equation (\ref{eq:me}) with a protocol $\l(\tau)$ that depends on $y_1$. The second measurement yields the information
\beq
\i_2= \ln[p_2(n_2,t_2|y_2,y_1)/p_1(n_2,t_2|y_1)]
\ee
with the analogous definition of $ p_2(n_2,t_2|y_2,y_1)  $ and $n(t_2)\equiv n_2$. In this process with feedback, the total change in system entropy  becomes
\beq
\Delta s\sys = -\ln p_2(n_t,t|y_2,y_1) + \ln p_0(n_0) .
\ee 
The total change of excess entropy can be split into three contributions from the three intervals $i=\{0\leq\tau<t_1\}, ii = \{t_1\leq \tau<t_2\}$ and $iii =\{ t_2\leq\tau\leq t\}$ as
\beq 
\Delta s\ex= \Delta s\ex_{i} + \Delta s\ex_{ii} + \Delta s\ex_{iii}.
\label{eq:sex}
\ee
With $\i=\i_1+\i_2$ and by combining Eqs. (\ref{eq:ent}) and (\ref{eq:i1}-\ref{eq:sex}), the average on the left hand side of Eq. (\ref{eq3}) still conditioned on the result $y_2,y_1$ can be written as
\begin{align}
\left\langle \frac{1}{p_0(n_0)}\right.&e^{-\Delta s\ex_i}\frac{p(n_1,t_1)}{p_1(n_1,t_1|y_1)}e^{-\Delta s\ex_{ii}} \frac{p_1(n_2,t_2|y_1)}{p_2(n_2,t_2|y_2,y_1)}\nonumber \\
&~~~~~~~~~~~~~~~~~~~~~~~~\times\left.e^{-\Delta s\ex_{iii}}p_2(n_t,t|y_2,y_1) \right\rangle& \nonumber \\
=\sum_{m_1,m_2}&\underbrace{\mean{\frac{1}{p_0(n_0)}e^{-\Delta s\ex_i}|n_1=m_1}_i p(m_1,t_1)}_{\alpha}\nonumber\\ 
\times&\underbrace{\mean{e^{-\Delta s\ex_{ii}}|n_1=m_1,n_2=m_2}_{ii}p_1(m_2,t_2|y_1)}_{\beta} \nonumber\\ 
\times&\underbrace{\mean{e^{-\Delta s\ex_{iii}}p_2(n_t,t|y_2,y_1)|n_2=m_2}_{iii}}_{\gamma} =1 \label{eq:ave}
\end{align}
where we have introduced conditioned averages on the three intervals $i$, $ii$ and $iii$ (indicated by subscripts) which eliminate the explicit factors $p_1^{-1}(n_1,t_1|y_1)$ and $p_2^{-1}(n_2,t_2|y_2,y_1)$. The term denoted by $\alpha$ is $1$ for any $m_1$ due to Eq. \eqref{eq:int2}. Likewise, the subsequent summation of $\beta$ over $m_1$ is 1 due to Eq. \eqref{eq:int3} as is the remaining sum of $\gamma$ over $m_2$. It should be obvious that each further measurements will lead to another summation of $\beta$-type and hence Eq. (\ref{eq3}) is proven for any number $N$  of measurements. This proof shows that Eq. (\ref{eq3}) holds true even as an average conditioned on a fixed record $\{y_\alpha, 1\leq \alpha\leq N \}$. In practice, a further average over all possible results $\{y_\alpha\}$ will typically be used as we also will for the remaining discussion.

Applying Jensen's inequality to Eq. (\ref{eq3})  leads to
\beq
-\Delta S\tot\leq- Q\hk + \I
\label{eq:Stot}
\ee
for the ensemble averages. For isothermal processes, we now invoke the first law
\beq
W\in - W\out= \Delta E + Q\hk + Q\ex= \Delta E +\Delta S\tot-\Delta S\sys .
\label{eq:W}
\ee
The splitting of the total work $W$ into a driving one and an extracted one depends on the specific system and is not fixed by these thermodynamic considerations. If $N\gg1$ measurements are repeated at times separated by $t_m$, inserting (\ref{eq:W}) in (\ref{eq:Stot}) and then dividing by $N t_m$ leads to the bound (\ref{eq4}) since both $\Delta E$ and $\Delta S\sys$ are finite boundary terms which vanish in the limit of a steady state operation.

As an aside, we note that the same reasoning applies if we start with the  valid modification of Eq. (\ref{eq:int1}) where $\Delta s\ex$ is replaced  by $\Delta s\ex+\Delta s\hk$ \cite{seif05a}. As a result, one obtains a concise proof of the Sagawa-Ueda equality \cite{saga10} $\langle \exp [-(\Delta s\tot + \i)] \rangle =1$ which is thus shown to hold true even for feedback-driven transitions between the genuine non-equilibrium states considered here. The resulting bound (\ref{eq2}), is,  of course, much weaker than (\ref{eq:Stot}) since the housekeeping heat is extensive in time.

{\sl Concluding perspective.--}
The equality \eqref{eq3} and the resulting inequalities \eqref{eq4} and \eqref{eq:Stot} generalize and sharpen previous results for feedback-driven systems with time-dependent Hamiltonians or potentials to transitions between genuine non-equilibrium states. Searching for the optimal protocol following a measurement is an issue that so far has primarily been asked for the former systems. It should now be investigated within this much wider class encompassing natural and artificial molecular machines driven by (or delivering output from) biochemical reactions. Our example of a simple driven two-state ``information machine'' has shown that its power becomes maximal if the time between measurements is the smallest and that it can even exceed the cost of driving. We suspect that both statements have a broad range of validity but for exploring its limits more analytical and numerical work will be necessary. Finally, the biggest challenge might be to come up with a first experimental demonstration of these relations. Whether the paradigmatic colloidal particle driven along a periodic potential or a molecular motor is more suitable for such an experiment remains to be seen.

\bibliography{feedback2}

%Merlin.mbs v4.21 2009-07-09.
\begin{thebibliography}{10}%
\makeatletter
\providecommand \@ifxundefined [1]{%
 \ifx #1\undefined \expandafter \@firstoftwo
 \else \expandafter \@secondoftwo
\fi
}%
\providecommand \@ifnum [1]{%
 \ifnum #1\expandafter \@firstoftwo
 \else \expandafter \@secondoftwo
\fi
}%
\providecommand \enquote [1]{``#1''}%
\providecommand \bibnamefont  [1]{#1}%
\providecommand \bibfnamefont [1]{#1}%
\providecommand \citenamefont [1]{#1}%
\providecommand\href[0]{\@sanitize\@href}%
\providecommand\@href[1]{\endgroup\@@startlink{#1}\endgroup\@@href}%
\providecommand\@@href[1]{#1\@@endlink}%
\providecommand \@sanitize [0]{\begingroup\catcode`\&12\catcode`\#12\relax}%
\@ifxundefined \pdfoutput {\@firstoftwo}{%
 \@ifnum{\z@=\pdfoutput}{\@firstoftwo}{\@secondoftwo}%
}{%
 \providecommand\@@startlink[1]{\leavevmode\special{html:<a href="#1">}}%
 \providecommand\@@endlink[0]{\special{html:</a>}}%
}{%
 \providecommand\@@startlink[1]{%
  \leavevmode
  \pdfstartlink
   attr{/Border[0 0 1 ]/H/I/C[0 1 1]}%
   user{/Subtype/Link/A<</Type/Action/S/URI/URI(#1)>>}%
  \relax
 }%
 \providecommand\@@endlink[0]{\pdfendlink}%
}%
\providecommand \url  [0]{\begingroup\@sanitize \@url }%
\providecommand \@url [1]{\endgroup\@href {#1}{\urlprefix}}%
\providecommand \urlprefix [0]{URL }%
\providecommand \Eprint[0]{\href }%
\@ifxundefined \urlstyle {%
  \providecommand \doi [1]{doi:\discretionary{}{}{}#1}%
}{%
  \providecommand \doi [0]{doi:\discretionary{}{}{}\begingroup
  \urlstyle{rm}\Url }%
}%
\providecommand \doibase [0]{http://dx.doi.org/}%
\providecommand \Doi[1]{\href{\doibase#1}}%
\providecommand \bibAnnote [3]{%
  \BibitemShut{#1}%
  \begin{quotation}\noindent
    \textsc{Key:}\ #2\\\textsc{Annotation:}\ #3%
  \end{quotation}%
}%
\providecommand \bibAnnoteFile [2]{%
  \IfFileExists{#2}{\bibAnnote {#1} {#2} {\input{#2}}}{}%
}%
\providecommand \typeout [0]{\immediate \write \m@ne }%
\providecommand \selectlanguage [0]{\@gobble}%
\providecommand \bibinfo [0]{\@secondoftwo}%
\providecommand \bibfield [0]{\@secondoftwo}%
\providecommand \translation [1]{[#1]}%
\providecommand \BibitemOpen[0]{}%
\providecommand \bibitemStop [0]{}%
\providecommand \bibitemNoStop [0]{.\EOS\space}%
\providecommand \EOS [0]{\spacefactor3000\relax}%
\providecommand \BibitemShut [1]{\csname bibitem#1\endcsname}%
%</preamble>
\bibitem{kawa07}%
  \BibitemOpen
  \bibfield{author}{%
  \bibinfo {author} {\bibfnamefont{R.}~\bibnamefont{Kawai}}, \bibinfo {author}
  {\bibfnamefont{J.~M.~R.}\ \bibnamefont{Parrondo}},\ and\ \bibinfo {author}
  {\bibfnamefont{C.}~\bibnamefont{{V}an~den Broeck}},\ }%
  \bibfield{journal}{%
  \bibinfo {journal} {Phys. Rev. Lett.}\ }%
  \textbf{\bibinfo {volume} {98}},\ \bibinfo {pages} {080602} (\bibinfo {year}
  {2007})%
  \bibAnnoteFile{NoStop}{kawa07}%
\bibitem{cao09}%
  \BibitemOpen
  \bibfield{author}{%
  \bibinfo {author} {\bibfnamefont{F.~J.}\ \bibnamefont{Cao}}\ and\ \bibinfo
  {author} {\bibfnamefont{M.}~\bibnamefont{Feito}},\ }%
  \bibfield{journal}{%
  \bibinfo {journal} {Phys. Rev. E}\ }%
  \textbf{\bibinfo {volume} {79}},\ \bibinfo {pages} {041118} (\bibinfo {year}
  {2009})%
  \bibAnnoteFile{NoStop}{cao09}%
\bibitem{saga10}%
  \BibitemOpen
  \bibfield{author}{%
  \bibinfo {author} {\bibfnamefont{T.}~\bibnamefont{Sagawa}}\ and\ \bibinfo
  {author} {\bibfnamefont{M.}~\bibnamefont{Ueda}},\ }%
  \bibfield{journal}{%
  \bibinfo {journal} {Phys. Rev. Lett.}\ }%
  \textbf{\bibinfo {volume} {104}},\ \bibinfo {pages} {090602} (\bibinfo {year}
  {2010})%
  \bibAnnoteFile{NoStop}{saga10}%
\bibitem{horo10}%
  \BibitemOpen
  \bibfield{author}{%
  \bibinfo {author} {\bibfnamefont{J.~M.}\ \bibnamefont{Horowitz}}\ and\
  \bibinfo {author} {\bibfnamefont{S.}~\bibnamefont{Vaikuntanathan}},\ }%
  \bibfield{journal}{%
  \Doi{10.1103/PhysRevE.82.061120}{\bibinfo {journal} {Phys. Rev. E}}\ }%
  \textbf{\bibinfo {volume} {82}},\ \bibinfo {pages} {061120} (\bibinfo {year}
  {2010})%
  \bibAnnoteFile{NoStop}{horo10}%
\bibitem{ponm10}%
  \BibitemOpen
  \bibfield{author}{%
  \bibinfo {author} {\bibfnamefont{M.}~\bibnamefont{Ponmurugan}},\ }%
  \bibfield{journal}{%
  \bibinfo {journal} {Phys. Rev. E}\ }%
  \textbf{\bibinfo {volume} {82}},\ \bibinfo {pages} {031129} (\bibinfo {year}
  {2010})%
  \bibAnnoteFile{NoStop}{ponm10}%
\bibitem{saga11}%
  \BibitemOpen
  \bibfield{author}{%
  \bibinfo {author} {\bibfnamefont{T.}~\bibnamefont{Sagawa}}\ and\ \bibinfo
  {author} {\bibfnamefont{M.}~\bibnamefont{Ueda}},\ \bibinfo {pages}
  {arXiv:1105.3262}}%
   (\bibinfo {year} {2011})%
  \bibAnnoteFile{NoStop}{saga11}%
\bibitem{toya10a}%
  \BibitemOpen
  \bibfield{author}{%
  \bibinfo {author} {\bibfnamefont{S.}~\bibnamefont{Toyabe}}, \bibinfo {author}
  {\bibfnamefont{T.}~\bibnamefont{Sagawa}}, \bibinfo {author}
  {\bibfnamefont{M.}~\bibnamefont{Ueda}}, \bibinfo {author}
  {\bibfnamefont{E.}~\bibnamefont{Muneyuki}},\ and\ \bibinfo {author}
  {\bibfnamefont{M.}~\bibnamefont{Sano}},\ }%
  \bibfield{journal}{%
  \bibinfo {journal} {Nat. Phys.}\ }%
  \textbf{\bibinfo {volume} {6}},\ \bibinfo {pages} {988} (\bibinfo {year}
  {2010})%
  \bibAnnoteFile{NoStop}{toya10a}%
\bibitem{mara10}%
  \BibitemOpen
  \bibfield{author}{%
  \bibinfo {author} {\bibfnamefont{R.}~\bibnamefont{Marathe}}\ and\ \bibinfo
  {author} {\bibfnamefont{J.~M.~R.}\ \bibnamefont{Parrondo}},\ }%
  \bibfield{journal}{%
  \bibinfo {journal} {Phys. Rev. Lett.}\ }%
  \textbf{\bibinfo {volume} {104}},\ \bibinfo {pages} {245704} (\bibinfo {year}
  {2010})%
  \bibAnnoteFile{NoStop}{mara10}%
\bibitem{vaik11}%
  \BibitemOpen
  \bibfield{author}{%
  \bibinfo {author} {\bibfnamefont{S.}~\bibnamefont{Vaikuntanathan}}\ and\
  \bibinfo {author} {\bibfnamefont{C.}~\bibnamefont{Jarzynski}},\ }%
  \bibfield{journal}{%
  \Doi{10.1103/PhysRevE.83.061120}{\bibinfo {journal} {Phys. Rev. E}}\ }%
  \textbf{\bibinfo {volume} {83}},\ \bibinfo {pages} {061120} (\bibinfo {year}
  {2011})%
  \bibAnnoteFile{NoStop}{vaik11}%
\bibitem{kim04}%
  \BibitemOpen
  \bibfield{author}{%
  \bibinfo {author} {\bibfnamefont{K.-H.}\ \bibnamefont{Kim}}\ and\ \bibinfo
  {author} {\bibfnamefont{H.}~\bibnamefont{Qian}},\ }%
  \bibfield{journal}{%
  \Doi{10.1103/PhysRevLett.93.120602}{\bibinfo {journal} {Phys. Rev. Lett.}}\
  }%
  \textbf{\bibinfo {volume} {93}},\ \bibinfo {pages} {120602} (\bibinfo {year}
  {2004})%
  \bibAnnoteFile{NoStop}{kim04}%
\bibitem{fuji10}%
  \BibitemOpen
  \bibfield{author}{%
  \bibinfo {author} {\bibfnamefont{Y.}~\bibnamefont{Fujitani}}\ and\ \bibinfo
  {author} {\bibfnamefont{H.}~\bibnamefont{Suzuki}},\ }%
  \bibfield{journal}{%
  \Doi{10.1143/JPSJ.79.104003}{\bibinfo {journal} {J. Phys. Soc. Jpn.}}\ }%
  \textbf{\bibinfo {volume} {79}},\ \bibinfo {pages} {104003} (\bibinfo {year}
  {2010})%
  \bibAnnoteFile{NoStop}{fuji10}%
\bibitem{horo11}%
  \BibitemOpen
  \bibfield{author}{%
  \bibinfo {author} {\bibfnamefont{J.~M.}\ \bibnamefont{Horowitz}}\ and\
  \bibinfo {author} {\bibfnamefont{J.~M.~R.}\ \bibnamefont{Parrondo}},\ }%
  \bibfield{journal}{%
  \Doi{10.1209/0295-5075/95/10005}{\bibinfo {journal} {EPL}}\ }%
  \textbf{\bibinfo {volume} {95}},\ \bibinfo {pages} {10005} (\bibinfo {year}
  {2011})%
  \bibAnnoteFile{NoStop}{horo11}%
\bibitem{abre11}%
  \BibitemOpen
  \bibfield{author}{%
  \bibinfo {author} {\bibfnamefont{D.}~\bibnamefont{Abreu}}\ and\ \bibinfo
  {author} {\bibfnamefont{U.}~\bibnamefont{Seifert}},\ }%
  \bibfield{journal}{%
  \Doi{10.1209/0295-5075/94/10001}{\bibinfo {journal} {EPL}}\ }%
  \textbf{\bibinfo {volume} {94}},\ \bibinfo {pages} {10001} (\bibinfo {year}
  {2011})%
  \bibAnnoteFile{NoStop}{abre11}%
\bibitem{hata01}%
  \BibitemOpen
  \bibfield{author}{%
  \bibinfo {author} {\bibfnamefont{T.}~\bibnamefont{Hatano}}\ and\ \bibinfo
  {author} {\bibfnamefont{S.~I.}\ \bibnamefont{Sasa}},\ }%
  \bibfield{journal}{%
  \Doi{10.1103/PhysRevLett.86.3463}{\bibinfo {journal} {Phys. Rev. Lett.}}\ }%
  \textbf{\bibinfo {volume} {86}},\ \bibinfo {pages} {3463} (\bibinfo {year}
  {2001})%
  \bibAnnoteFile{NoStop}{hata01}%
\bibitem{toya10}%
  \BibitemOpen
  \bibfield{author}{%
  \bibinfo {author} {\bibfnamefont{S.}~\bibnamefont{Toyabe}}, \bibinfo {author}
  {\bibfnamefont{T.}~\bibnamefont{Okamoto}}, \bibinfo {author}
  {\bibfnamefont{T.}~\bibnamefont{Watanabe-Nakayama}}, \bibinfo {author}
  {\bibfnamefont{H.}~\bibnamefont{Taketani}}, \bibinfo {author}
  {\bibfnamefont{S.}~\bibnamefont{Kudo}},\ and\ \bibinfo {author}
  {\bibfnamefont{E.}~\bibnamefont{Muneyuki}},\ }%
  \bibfield{journal}{%
  \Doi{10.1103/PhysRevLett.104.198103}{\bibinfo {journal} {Phys. Rev. Lett.}}\
  }%
  \textbf{\bibinfo {volume} {104}},\ \bibinfo {pages} {198103} (\bibinfo {year}
  {2010})%
  \bibAnnoteFile{NoStop}{toya10}%
\bibitem{haya10}%
  \BibitemOpen
  \bibfield{author}{%
  \bibinfo {author} {\bibfnamefont{K.}~\bibnamefont{Hayashi}}, \bibinfo
  {author} {\bibfnamefont{H.}~\bibnamefont{Ueno}}, \bibinfo {author}
  {\bibfnamefont{R.}~\bibnamefont{Iino}},\ and\ \bibinfo {author}
  {\bibfnamefont{H.}~\bibnamefont{Noji}},\ }%
  \bibfield{journal}{%
  \Doi{10.1103/PhysRevLett.104.218103}{\bibinfo {journal} {Phys. Rev. Lett.}}\
  }%
  \textbf{\bibinfo {volume} {104}},\ \bibinfo {pages} {218103} (\bibinfo {year}
  {2010})%
  \bibAnnoteFile{NoStop}{haya10}%
\bibitem{blic07}%
  \BibitemOpen
  \bibfield{author}{%
  \bibinfo {author} {\bibfnamefont{V.}~\bibnamefont{Blickle}}, \bibinfo
  {author} {\bibfnamefont{T.}~\bibnamefont{Speck}}, \bibinfo {author}
  {\bibfnamefont{C.}~\bibnamefont{Lutz}}, \bibinfo {author}
  {\bibfnamefont{U.}~\bibnamefont{Seifert}},\ and\ \bibinfo {author}
  {\bibfnamefont{C.}~\bibnamefont{Bechinger}},\ }%
  \bibfield{journal}{%
  \Doi{10.1103/PhysRevLett.98.210601}{\bibinfo {journal} {Phys. Rev. Lett.}}\
  }%
  \textbf{\bibinfo {volume} {98}},\ \bibinfo {pages} {210601} (\bibinfo {year}
  {2007})%
  \bibAnnoteFile{NoStop}{blic07}%
\bibitem{gome09}%
  \BibitemOpen
  \bibfield{author}{%
  \bibinfo {author} {\bibfnamefont{J.~R.}\ \bibnamefont{Gomez-Solano}},
  \bibinfo {author} {\bibfnamefont{A.}~\bibnamefont{Petrosyan}}, \bibinfo
  {author} {\bibfnamefont{S.}~\bibnamefont{Ciliberto}}, \bibinfo {author}
  {\bibfnamefont{R.}~\bibnamefont{Chetrite}},\ and\ \bibinfo {author}
  {\bibfnamefont{K.}~\bibnamefont{Gawedzki}},\ }%
  \bibfield{journal}{%
  \Doi{10.1103/PhysRevLett.103.040601}{\bibinfo {journal} {Phys. Rev. Lett.}}\
  }%
  \textbf{\bibinfo {volume} {103}},\ \bibinfo {pages} {040601} (\bibinfo {year}
  {2009})%
  \bibAnnoteFile{NoStop}{gome09}%
\bibitem{jarz97}%
  \BibitemOpen
  \bibfield{author}{%
  \bibinfo {author} {\bibfnamefont{C.}~\bibnamefont{Jarzynski}},\ }%
  \bibfield{journal}{%
  \bibinfo {journal} {Phys. Rev. Lett.}\ }%
  \textbf{\bibinfo {volume} {78}},\ \bibinfo {pages} {2690} (\bibinfo {year}
  {1997})%
  \bibAnnoteFile{NoStop}{jarz97}%
\bibitem{seif10}%
  \BibitemOpen
  \bibfield{author}{%
  \bibinfo {author} {\bibfnamefont{U.}~\bibnamefont{Seifert}},\ }%
  \bibfield{journal}{%
  \Doi{10.1103/PhysRevLett.104.138101}{\bibinfo {journal} {Phys. Rev. Lett.}}\
  }%
  \textbf{\bibinfo {volume} {104}},\ \bibinfo {pages} {138101} (\bibinfo {year}
  {2010})%
  \bibAnnoteFile{NoStop}{seif10}%
\bibitem{seif05a}%
  \BibitemOpen
  \bibfield{author}{%
  \bibinfo {author} {\bibfnamefont{U.}~\bibnamefont{Seifert}},\ }%
  \bibfield{journal}{%
  \bibinfo {journal} {Phys. Rev. Lett.}\ }%
  \textbf{\bibinfo {volume} {95}},\ \bibinfo {pages} {040602} (\bibinfo {year}
  {2005})%
  \bibAnnoteFile{NoStop}{seif05a}%
\end{thebibliography}%

\end{document}